\documentstyle[12pt,epsf]{article}

\newcommand{\tfrac}[2]{{\textstyle\frac{#1}{#2}}}
\begin{document}
\begin{titlepage}
\title{New Universality Class in the Superconductive Phase Transition
\thanks{Invited talk presented by A. Schakel at the 10'th International
Conference on Problems of Quantum Field Theory, Alushta, Ukraine, May 13-17,
1996, based on work carried out in collaboration with M. Kiometzis
\protect\cite{berlin}.}}  \author{Hagen Kleinert\thanks{Email:
kleinert@physik.fu-berlin.de.  Home page:
http://www.physik.fu-berlin.de/$\sim$kleinert} \, and Adriaan
M. J. Schakel\thanks{Email: schakel@physik.fu-berlin.de} \\ Institut f\"ur
Theoretische Physik \\ Freie Universit\"at Berlin \\ Arnimalee 14, 14195
Berlin.}  \date{June 12, 1996} \maketitle
\begin{abstract}
The superconductive phase transition (or Coulomb-Higgs phase transition of scalar QED in 3D)
is discussed in a dual
formulation which focuses on the magnetic rather than the electric excitations
of the system.  Renormalization group analysis of the dual formulation reveals
the transition to be of second order and of a new universality class. Whereas
coherence length and specific heat have $XY$-model exponents, the magnetic
penetration depth shows mean-field behavior.  Experimental evidence for these
predictions is discussed.
\end{abstract}
\end{titlepage}
The study of
Theoretical studies of the superconductive phase transition 
in the Ginzburg-Landau theory (or Coulomb-Higgs phase transition of scalar QED in 3D)
have a long history. In the the early seventies, Halperin,
Lubensky, and Ma \cite{HLM} analyzed the transition in a on-loop
$\epsilon$-expansion and noted the absence of an infrared stable (IR) fixed
point, suggesting that the transition be first order.  A similar conclusion
had been reached earlier in 4D by Coleman and Weinberg \cite{CW}.

The Lagrange density describing scalar QED reads, in the Euclidean formulation,
\begin{equation}  \label{L}
{\cal L} = \tfrac{1}{4} (\partial_\mu A_\nu - \partial_\nu A_\mu)^2 +
|(\partial_\mu -i q A_\mu)\phi|^2 + m_\phi^2 |\phi|^2 + \lambda |\phi|^4 ,
\end{equation}
where $A_\mu$ is the electromagnetic gauge field, and $\phi$ the
self-interacting order or Higgs field with charge $q>0$, mass $m_\phi$ and coupling
constant $\lambda$.  As in superconductivity, the mass term is assumed to
depend on some external parameter such as temperature $T$, changing sign at
some critical value $T_{\rm c}$.  Thus we set $m_\phi^2 \equiv \xi_0^{-2}\tau
\equiv \xi_0^{-2}(T/T_{\rm c}-1)$, where $\xi_0$ is the length scale of the
amplitude fluctuations of the Higgs field.

\begin{figure}
\vspace{-1.6cm}
\epsfxsize=7.cm
\begin{center}
\mbox{\epsfbox{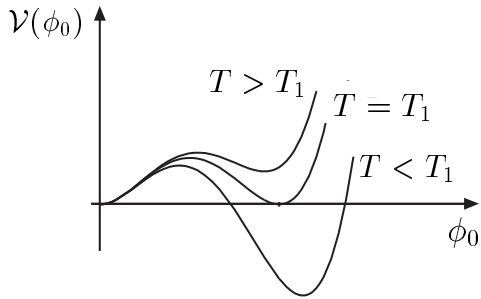}}
\end{center}
\vspace{-0.8cm}
\caption{Effective potential up to one-loop order.}
\label{fig:pot}
\vspace{-.4cm}
\end{figure}

At the mean-field level, the theory undergoes a transition at $T=T_{\rm c}$.
The high-temperature Coulomb phase is characterized by a zero average of the
Higgs field, while in the low-temperature Higgs phase this average is
non-zero.  For definiteness we consider the London limit, where the Higgs
field is assumed to have a constant amplitude $\phi(x) =
\frac{1}{\sqrt{2}}\phi_0 \exp[i\theta(x)]$.  The integration over the phase
field $\theta$ is easily carried out and results in a mass term $\frac{1}{2}
m_A^2 A_\mu^2$ for the gauge field, with $m^2_A = q^2 \phi_0^2$, if we work in
the Lorentz gauge $\partial_\mu A_\mu=0$.  If we integrate out also the gauge
field in the partition function, we obtain a one-loop effective potential
\begin{equation} \label{Veff}
{\cal V}_{\rm eff}(\phi_0)  = \int_{k<\Lambda } \frac{{\rm d}^3k}{(2 \pi)^3}
\ln(k^2 + m_A^2) =  - \frac{1}{6 \pi}  q^3 \phi_0^3.
\end{equation}
The integral is regularized by an UV cutoff $\Lambda$, and the UV divergences
of the type $\Lambda^3\log \Lambda$ and $\Lambda \phi_0^2$ are absorbed
into the total energy and mass, as usual.  The potential ${\cal V}_{\rm
eff}(\phi_0) $ adds to the tree potential ${\cal V}_0 = \frac{1}{2}m_\phi^2
\phi_0^2 + \frac{1}{4}\lambda \phi_0^4$, and results in a first-order phase
transition {\em above\/} $T_{\rm c}$ at $T_1 = T_{\rm c} (1+\xi_0^2 q^6/18
\pi^2 \lambda)$, see Fig.\ \ref{fig:pot}.

This conclusion has been questioned on various grounds.  First of all, there
is the experimental accessible smectic-A to nematic phase transition in liquid
crystals.  De Gennes' theory describing this transition can be mapped onto a
Ginzburg-Landau model with the transverse fluctuations of the director field
in the smectic-A phase playing the role of the gauge field \cite{HL}.  The
transition is known to have a second-order regime \cite{thoen}.  Furthermore,
a lattice simulation carried out by Dasgupta and Halperin \cite{DH}, employing
a duality map introduced by Peskin, Thomas and Stone \cite{PTS}, showed that
for small $q$-values the theory undergoes a second-order transition.

To see the flaw in the argument leading to the previous conclusion, consider
the Ginzburg region $|\tau_{\rm G}|< \lambda^2 \xi_0^2$ around $T_{\rm c}$ in
which amplitude fluctuations of the Higgs field can no longer be ignored.
There, the London limit breaks down and the amplitude fluctuations of the
Higgs field must be included.  They have a mass term $\frac{1}{2} m^2_{\rm H}
{\tilde \phi}^2$ with $m_{\rm H} = \sqrt{2} |m_\phi|$ the so-called Higgs
mass.  The ratio of the Ginzburg region to the interval $\tau_1=T_1/T_{\rm
c}-1$ in which the first-order jump takes place $\tau_{\rm G}/\tau_1 \sim
\lambda^3/q^6 \sim \kappa^6$ is proportional to the sixth power of the
Ginzburg-Landau parameter $\kappa = m_{\rm H}/\sqrt{2}m_A$.  As has
been stressed by Kleinert \cite{tricritical}, in type-II superconductors where
$\kappa > 1/\sqrt{2}$, the first-order jump takes place inside the Ginzburg
region, where the amplitude fluctuations around $\phi_0$ are no longer smooth
and of low energy, but driven by magnetic vortex loops.  At the core of a
vortex, the Higgs field vanishes making the effective potential (\ref{Veff})
unreliable.  As will be seen below, these fluctuations increase with
increasing $\kappa$.

To describe a vortex loop, it is useful to first construct a vortex line with
a source at some space-time point $z$, which acts like magnetic monopole.  As
such it is accompanied by a Dirac string.  The electrodynamics in the presence
of a monopole was first described by Dirac \cite{Dirac} who argued that
$F_{\mu \nu} = \partial_\mu A_\nu - \partial_\nu A_\mu - F_{\mu \nu}^{\rm P}$
plays the role of a physical field strength. The subtracted term
\begin{equation}
F_{\mu \nu}^{\rm P}(x) = \epsilon_{\mu \nu \lambda} \Phi_0 \int_{L_z} {\rm d}
y_\lambda \, \delta(x-y) = \epsilon_{\mu \nu \lambda} \Phi_0 \int_{L_z} {\rm
d} s \frac{{\rm d} y_\lambda }{{\rm d} s} \delta[x-y(s)],
\end{equation}
with $\Phi_0 = 2 \pi/q$ being the magnetic flux quantum, removes the field of
the Dirac string running along some path $L_z$ from the location $z$ of the
monopole to infinity.

The field $F_{\mu \nu}^{\rm P}(x)$ is a gauge field in its own right, with the
corresponding gauge transformations changing the string configuration
\cite{strigf}.

The monopole density is given by the divergence of the dual field ${\tilde
F}_{\mu}^{\rm P} = \frac{1}{2} \epsilon_{\mu \nu \lambda} F_{\nu \lambda
}^{\rm P}$, i.e., $\partial_\mu {\tilde F}^{\rm P}_\mu (x) = \Phi_0 \, \delta
(x - z)$.  In the Higgs phase, the Dirac string becomes the core of a vortex
\cite{Nambu}.

The field equation obtained from the $A_\mu$-dependent terms in the Lagrange
density, ${\cal L}_A = \frac{1}{2} \tilde{F}_\mu^2 + \frac{1}{2} m_A^2
A_\mu^2$, is solved by
\begin{equation}
\tilde{F}_\mu(x) = -m_A^2 \int {\rm d}^3 y \, G(x-y) \tilde{F}_\mu^{\rm P}(y)
+ \Phi_0 \partial_\mu G(x-z); \;\;\;
G(x) = \int \frac{{\rm d}^3k}{(2 \pi)^3} \frac{{\rm e}^{ik.x}}{k^2 + m_A^2},
\end{equation}
with $G(x)$ the massive propagator.  Substituting the solution back into the
Lagrange density, we obtain, ignoring the (diverging) monopole
self-interaction, the action $S_A = M_V \int_{L_z} {\rm d} s$ of a
relativistic point particle with mass
\begin{equation} \label{M_V}
M_V = \frac{1}{2} m_A^2 \Phi_0^2 \int \frac{{\rm d}^2 k}{(2
\pi)^2} \frac{1}{{\bf k}^2 + m_A^2}
\approx \frac{1}{4 \pi}m_A^2 \Phi_0^2 \ln \kappa.
\end{equation}
In deriving this we employed the Lorentz invariance of the theory which
allowed us to calculate the mass in the rest frame of the particle, and we
took a vortex radius of the order of the coherence length $1/|m_\phi|$ .  We
see that in the Higgs phase, the vortex surrounding the Dirac string acts as
the worldline of a point particle.  Equation (\ref{M_V}) was first obtained by
Abrikosov in the context of superconductivity \cite{Abrikosov}.  To see in
which regime these excitations are important, we convert the mass $M_V$ into a
magnetic field via $H_{{\rm c}_1} = M_V/\Phi_0$, and compare it with the
critical field $H_{\rm c} = \Phi_0 |m_\phi| m_A/2\sqrt{2} \pi$ which is
obtained by equating the tree potential ${\cal V}_0$ to $\frac{1}{2} H_{\rm
c}^2$.  This field is a measure of the condensation energy which in turn sets
the energy scale.  It physically denotes the external field at which a type-I
system can no longer resist the magnetic pressure and reverts to the Coulomb
phase characterized by a perfect penetration of the field.  The ratio of the
two fields $H_{{\rm c}_1}/H_{\rm c} = \ln \kappa/\, {\sqrt{2} \kappa}$ shows
that for increasing $\kappa$, vortices become easier to excite.

The field $V(L_z)$ describing the vortex is easily inferred by noting that in
the functional-integral approach, a given field configuration is weighed by a
factor ${\rm e}^{-S}$.  The $\tilde{F}_\mu^{\rm P}$-dependent terms then
defines the vortex field \cite{Ezawa}: $V(L_{z}) = \exp [ \int {\rm d}^3 x
(\tfrac{1}{2} {\tilde F}_\mu^{{\rm P}^2} + {\tilde F}_\mu {\tilde F}_\mu^{\rm
P})]$.  The Poisson bracket $\{F_{0i}(t,{\bf x}), A_j(t,{\bf y})\} = -
\delta_{ij} \delta({\bf x}-{\bf y})$, stating that $F_{0i}$ is canonical
conjugate to $A_i$, implies that the vortex field and, thus, the point
particle carries one flux quantum as charge: $\{V,\Phi\} = \Phi_0 V$.  We
shall henceforth refer to the point particles as fluxons.  The symmetry
generated by the magnetic flux $\Phi = \int {\rm d}^2 x {\tilde F}_0$ is a
global U(1) symmetry which is to be distinguished from the local U(1) gauge
symmetry.  As has been pointed out by Kovner, Rosenstein and collaborators
\cite{KR}, although the flux symmetry is of topological nature, the
corresponding current ${\tilde F}_\mu$ being trivially conserved when
expressed in terms of the gauge field, it is apparently not devoid of physical
relevance.  The central point of the dual formulation of the Abelian Higgs
model is to describe the fluxons by a field theory where the flux symmetry
becomes an ordinary Noether symmetry with a current whose conservation follows
upon invoking the field equations.

Let us proceed to calculate the correlation function $\langle V( L_{{z}} )
V^*( L_{\bar {z}} ) \rangle$ where $V^*( L_{\bar {z}} )$ describes an
antimonopole located at ${\bar z}$ and its accompanying Dirac string.  In the
Lorentz gauge, we have
\begin{equation} \label{init}
\langle V( L_{z} ) V^*(L_{\bar {z}}) \rangle = \int {\cal D} A_\mu \exp\left[-
\tfrac{1}{2} \int {\rm d}^3 x \left({\tilde F}_\mu^2 + m_A^2 A_\mu^2
\right)\right]
\end{equation}
where now $\partial_\mu {\tilde F}_\mu (x) = \rho(x)$, with monopole density
$\rho(x) = \Phi_0 [ \delta (x-z)- \delta(x - {\bar z})]$.  Equation
(\ref{init}) can be evaluated directly by integrating out the gauge field.
For later convenience, however, we first linearize the Maxwell term via a
Hubbard-Stratonovich transformation so that $-\frac{1}{2} {\tilde F}_\mu^2
\rightarrow - \frac{1}{2} h_\mu^2 + i h_\mu {\tilde F}_\mu$.  The physical
relevance of the auxiliary field becomes apparent when considering its field
equation, $h_\mu = i {\tilde F}_\mu$, showing that it represents the dual
field.  In this case, where the dual field is described directly by $h_\mu$
and not via an unphysical gauge potential, the flux symmetry is no longer a
topological symmetry.  The correlation function becomes
\begin{equation} \label{correl}
\langle V ( L_z )  V^*(
L_{\bar z}) \rangle = \int {\cal D} h_\mu \,
\exp \left\{- \int {\rm d}^3 x \left[\frac{1}{4m_A^2}(\partial_\mu h_\nu -
\partial_\nu h_\mu)^2 +\frac{1}{2}  h_\mu^2   + i h_\mu {\tilde F}^{\rm
P}_\mu \right] \right\},
\end{equation}
showing that the magnetic vortex couples with a charge $g=\Phi_0 m_A$ to the
massive vector field $h_\mu$ (the factor $\Phi_0$ is hidden in the definition
of the dual field ${\tilde F}^{\rm P}_\mu$, while the mass factor arises after
rescaling the vector field so that the kinetic term has the usual coefficient
${1}/{4}$).  The remaining integral in (\ref{correl}) is Gaussian and can be
easily carried out to yield:
\begin{equation}
\langle V( L_{{z}} ) V^*( L_{\bar {z}} )
\rangle =  \exp\left(\frac{\pi}{q^2} \frac{
{\rm e}^{-m_A\left| L_{{{z}}{\bar {z}}} \right|} }{ \left|
L_{{{z}}{\bar {z}}} \right| } \right) \, {\rm e}^{-M_V\left|
L_{{{z}}{\bar {z}}} \right|} .
\end{equation}
In the Higgs phase, where the fluxon mass is nonzero, the second factor
dominates and the correlation function tends to zero in the limit where the
monopole-antimonopole pair is taken very far apart, so that the length of the
vortex connecting the pair $| L_{{{z}}{\bar {z}}} |$ tends to infinity.  This
physically represents monopole confinement.  In the Coulomb phase, where $m_A$
and therefore the fluxon mass is zero, we have on the other hand $\langle V(
L_{z} ) V^*( L_{\bar {z}}) \rangle \rightarrow 1$.  On account of the cluster
property of correlation functions, this implies that the vortex field develops
a vacuum expectation value, thus signaling a proliferation of magnetic
vortices.  It should be noted that it is the Coulomb phase and not the Higgs
phase where $V(L_z)$ develops an expectation value.  It follows that the
vortex field can be used to distinguish the two phases, i.e., it is an order
parameter---or better, a disorder parameter \cite{GFCM}.

It is instructive to consider the limit $m_A \rightarrow 0$ directly in
(\ref{correl}).  Because of its diverging coefficient, the first term must
vanish, i.e., we have $h_\mu = \partial_\mu \gamma$, and
\begin{eqnarray} \label{VV}
\langle V( L_z) V^*(L_{\bar z}) \rangle = \int {\cal D} \gamma \exp \left\{
\int
{\rm d}^3 x \left[ -\tfrac{1}{2} (\partial_\mu \gamma)^2 + i \gamma \rho
\right]
\right\}.
\end{eqnarray}
This shows that in the Coulomb phase, $V(L_z)$ is represented as a normal
field $V( L_z) = \exp[i \Phi_0 \gamma (z)]$ and not via a singular field
${\tilde F}^{\rm P}_\mu$.  Apparently, the Dirac string has become irrelevant
here.  The massless scalar field appearing in (\ref{VV}) is the Goldstone
field of the broken global U(1) flux symmetry.  The corresponding current
$j_\mu = \partial_\mu \gamma$ is no longer trivially conserved, but rather has
become a Noether current whose conservation follows upon employing the field
equation $\partial^2 \gamma=0$.  Physically, $\gamma$ represents the massless
photon of the Coulomb phase.  Since in two space and one time dimension the
photon has only one transverse direction and thus only one degree of freedom,
it can be represented by a scalar field.

We are now in a position to write down the dual theory which physically
describes a grand canonical ensemble of fluctuating closed magnetic vortices,
of arbitrary length and shape.  We will refer to such an ensemble as a loop
gas.  A neutral loop gas can be described by a massive scalar field theory
with a repulsive two-particle interaction $u |\psi|^4$, accounting for the
steric repulsion between vortices.  In our case, the vortices interact in
addition with the massive vector field $h_\mu$, with coupling constant
$g=\Phi_0 m_A$.  This gives as dual theory \cite{GFCM,KR,berlin}
\begin{equation} \label{Ltilde}
\tilde{\cal L} = \frac{1}{4 m_A^2} (\partial_\mu h_\nu - \partial_\nu h_\mu)^2
+ \frac{1}{2} h_\mu^2 + |(\partial_\mu -i \Phi_0 h_\mu)\psi|^2 + m_\psi^2
|\psi|^2 + u |\psi|^4 ,
\end{equation}
where the $\psi$-field is a {\em disorder field\/} whose Feynman diagrams are
direct spacetime pictures of the fluctuating vortex loops in the original
system (\ref{L}) \cite{GFCM}.  It is minimally coupled to the magnetic field
$h_\mu$.  The dual theory describes the system directly in terms of physical
degrees of freedom; there is no local gauge symmetry.  The old mass result
(\ref{M_V}) appears in the dual theory as a one-loop contribution to the
self-energy of the scalar field $\psi$.  The interaction strength of $h_\mu$
is determined by the dimensionless ratio $g^2/m_A$, which is the inverse of
that of the gauge field, $g^2/m_A \sim m_A/q^2$.  This is a common feature of
mutually dual theories.  It is one of the reasons why the dual theory is more
easily dealt with than the original Higgs model, where there is no IR fixed
point in the weak-coupling regime, leaving only the strong-coupling regime for
a second-order phase transition.

Before carrying out the renormalization program we note that in the dual
theory no cubic term is generated when the vector field is integrated out
\cite{tricritical}.  More precisely, repeating the steps leading to the
effective potential ${\cal V}_{\rm eff} (\phi)$, Eq.\ (\ref{Veff}), we obtain
now
\begin{equation}
{\cal V}_{\rm eff}(\psi) =  \int_{k<\Lambda } \frac{{\rm d}^3k}{(2 \pi)^3}
\ln(k^2 + m_A^2 + 2 g^2 |\psi|^2) =  - \frac{1}{6 \pi}  (m_A^2 + 2 g^2
|\psi|^2)^{3/2}.\label{expan}
\end{equation}
This has a Tayler expansion in $|\psi|^2$ and, in contrast to the previous
case, no cubic term is generated in the Higgs phase characterized by a
non-zero vector field mass.  Since $g = \phi_0 m_A$, the effective potential
${\cal V}_{\rm eff}(\psi)$ vanishes as we approach the critical point by
letting $m_A \rightarrow 0$.  Thus the massive vector field decouples from the
$\psi$-field representing the vortices, and the Coulomb-Higgs phase transition
reduces to a pure $|\psi|^4$-theory, which is known to undergo a second-order
transition.

We now carry out the renormalization program, and write the bare theory
(\ref{Ltilde}) as a sum of the renormalized Lagrange density and counter terms
\begin{eqnarray}
\delta \tilde{\cal L} = \!\!\!\! && \!\!\!\! Z_\psi |(\partial_\mu -i g h_\mu
Z_g/Z_\psi ) \psi|^2 - |(\partial_\mu -i g h_\mu) \psi|^2 + (Z_\psi
m_{\psi,0}^2
- m_\psi^2) |\psi|^2 \nonumber \\ && \!\!\!\!+ u (Z_u-1) |\psi|^4 +
\tfrac{1}{4}(Z_{h} - 1) (\partial_\mu h_\nu - \partial_\nu h_\mu)^2 +
\tfrac{1}{2} (Z_{h} m_{A,0}^2- m_A^2) h_\mu^2 ,
\end{eqnarray}
where $h_\mu = Z_{h}^{-1/2} h_{0,\mu}, \, g = Z_g^{-1} Z_\psi Z_{h}^{1/2} g_0,
\, \psi = Z_{\psi}^{-1/2} \psi_0,$ and $u = Z_{u}^{-1} Z_{\psi}^2 \, u_0$.
[All quantities appearing in (\ref{Ltilde}) have been given an index 0 to
indicate that they refer to bare quantities.]  An explicit one-loop
calculation \cite{berlin}, which uses the fixed-dimension approach of Parisi
\cite{Parisi}, revealed that $Z_g = Z_\psi$, so that the minimal coupling is
preserved at the one-loop order, and also that the mass term did not receive a
correction, so that $m_{A} = Z_{h}^{1/2} m_{A,0}$.  Both these statements
remain true to all orders in perturbation theory on account of the Ward
identity known from QED which also operates in the massive vector case.  As a
result, both the mass and coupling constant of the dual theory renormalize in
the same way, and they are linked via the equation $g = (2\pi/q) m_A$, with
the electric charge $q$ a free parameter which does not renormalize in the
dual formulation.

\begin{figure}
\vspace{-.9cm}
\epsfysize=3.cm
\begin{center}
\mbox{\epsfbox{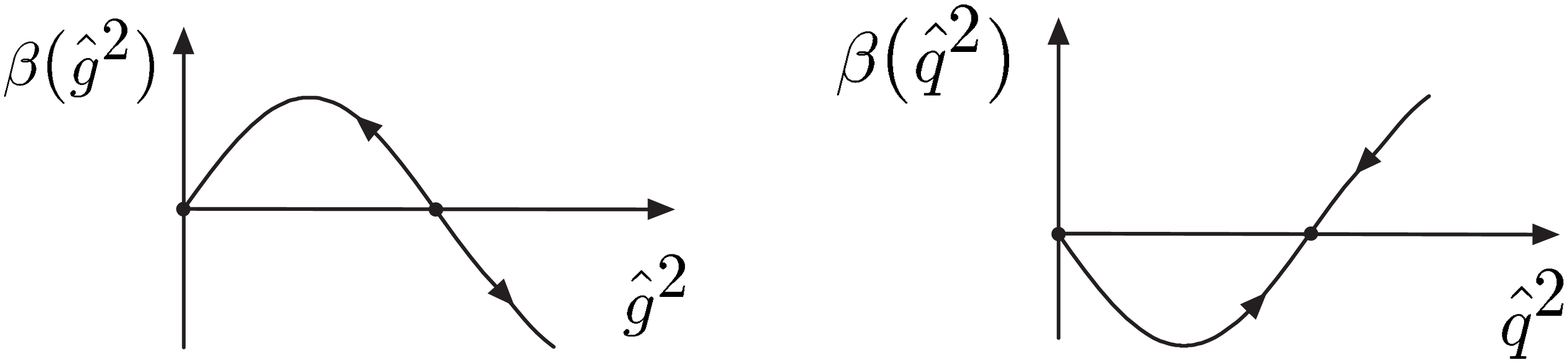}}
\end{center}
\vspace{-0.5cm}
\caption{The $\beta$-function of the dual theory (left figure), showing that
the origin is an IR fixed point.  For comparison, the expected form of the
$\beta$-function of the Higgs model (right figure) is included.}
\vspace{-0.4cm}
\label{fig:beta}
\end{figure}

The double role played by $m_A$ as both a mass parameter and a coupling
constant is essential to our renormalization group analysis \cite{berlin}.  We
assume, as suggested by the first-loop calculation, that the phase transition
is driven by the proliferation of vortices as signaled by a non-zero
expectation value of the $\psi$-field.  On account of the relation $g = \Phi_0
m_A$, we can cast the $\beta$-function $\beta_\psi (\hat{g}^2,\hat{u}) :=
\partial \hat{g}^2/\partial \ln(m_\psi) |_0$, with
$\hat{u}:=u/m_\psi$ and the index 0 indicating that $g_0$ and $u_0$ are kept
fixed, of the properly scaled coupling constant squared,
$\hat{g}^2 := g^2/m_\psi$, in the form
$$
\beta_\psi (\hat{g}^2,\hat{u}) = \hat{g}^2 \left[-1 + \frac{\partial \ln(g^2)
}{\partial \ln(m_\psi)}\right]\Biggr|_0 = \hat{g}^2 \left[-1 +
\frac{\partial \ln(m_{A,0}^2) }{\partial \ln(m_\psi)} + \frac{\partial
\ln(Z_h) }{\partial \ln(m_A)} \frac{\partial \ln(m_A) }{\partial \ln(m_\psi)}
\right] \Biggr|_0.
$$
Since $m_{A,0}^2 \sim m_{\psi,0}^2 \sim \tau$, the second term at the
right-hand side is the inverse of the correlation-length exponent $\nu$, which
indicates how fast the renormalized mass $m_\psi$ tends to zero, $m_\psi \sim
\tau^\nu$.  The last term contains the
$\gamma_{h}(\hat{g}^2,\hat{u})$-function, $\partial \ln(Z_h)/\partial
\ln(m_A)|_0$ which yields the anomalous dimension $\eta_h$ of the vector field
when evaluated at the fixed point.  That is,
\begin{equation} \label{beta}
\beta_\psi (\hat{g}^2,\hat{u}) = \hat{g}^2\left[-1 +\frac{1}{\nu} + \gamma_{
h}(\hat{g}^2,\hat{u}) \frac{\partial \ln(m_A)}{\partial \ln (m_\psi)}
\Biggr|_0 \right].
\end{equation}
Without the double role played by $m_A$, the coefficient of the $\hat{g}^2$
term in the $\beta_\psi (\hat{g}^2,\hat{u})$-function would be $-1$, implying
that the origin $\hat{g}=0$ is an UV fixed point.  Here, however, the
coefficient is $-1+1/\nu$ which is positive if $\nu < 1$.  In $3D$, where $\nu
\approx \frac{2}{3}$, this is the case, meaning that the origin $\hat{g}^2=0$
is IR stable and the massive vector field decouples, see Fig.\ \ref{fig:beta}.
Hence, we are left with the pure $|\psi|^4$-theory, just as we saw by
expanding (\ref{expan}).

This situation differs sharply from that in the original formulation, where
the coupling $q$ to the electromagnetic gauge field is assumed to possess a
perturbatively inaccessible IR fixed point away from the origin, see
Fig.\ \ref{fig:beta}.  The Gaussian fixed point $\hat{g}^*=\hat{u}^*=0$ is a
tricritical point of the Higgs model whose existence and location was first
established by Kleinert \cite{tricritical}.  At this point, where the
second-order transition changes into a first-order one, the vortices are
completely free.

We conclude that the Coulomb-Higgs phase transition constitutes a new
universality class.  Whereas the disorder field behaves as in a pure
$|\psi|^4$-theory, the exponents of the magnetic field retain their mean-field
values.

Up to recently, this problem was of mere academic interest because the
temperature resolution needed to probe the Ginzburg region could not be
realized experimentally.  However, in various high-$T_{\rm c}$ materials, this
region has become fairly large $|\tau_{\rm G}| \sim 10^{-2}$, and
experimentally accessible.  Experiments just carried out by Lin {\it et al.}
\cite{Lin} on YBaCuO seem to confirm our predictions.  They clearly show the
mean-field behavior of the magnetic penetration depth. Earlier measurements of
the specific heat had established the $XY$-model-like behavior of the heat
capacity \cite{xymod}.

\end{document}